# Empirical Models for the Realistic Generation of Cooperative Awareness Messages in Vehicular Networks

Rafael Molina-Masegosa, Miguel Sepulcre, Javier Gozalvez, Friedbert Berens, Vincent Martinez

*Abstract*—Most V2X (Vehicle-to-Everything) applications rely on broadcasting awareness messages known as CAM (Cooperative Awareness Messages) in ETSI or BSM (Basic Safety Message) in SAE standards. A large number of studies have been devoted to guarantee their reliable transmission. However, to date, the studies are generally based on simplified data traffic models that generate awareness messages at periodic intervals or with a constant message size. These models do not accurately represent the real generation of CAM messages that follow specific mobility-based rules. Using simplified and unrealistic traffic models can significantly impact the results and validity of the studies, and hence accurate models for the generation of awareness messages are necessary. This paper proposes the first set of models that can realistically generate CAM messages. The models have been created from real traces collected by two car manufacturers in urban, sub-urban and highway test drives. The models are based on $m$th order Markov sources, and model the size of CAMs and the time interval between CAMs. The models are openly provided to the community and can be easily integrated into any simulator.

## I. Introduction

V2X networks will support connected and automated driving thanks to the wireless exchange of information. Most V2X-enabled applications rely on frequently broadcasting awareness messages known as CAM (Cooperative Awareness Messages) or BSM (Basic Safety Message) in ETSI and SAE standards, respectively. These messages include the position, speed and basic status information of the transmitting node. These messages are independent of the underlying wireless technologies (e.g., ITS-G5, DSRC, LTE-V2X or 5G V2X). Their importance has triggered a large number of studies aimed at guaranteeing their reliable transmission. Most of these studies utilize simplified traffic models for the generation of the awareness messages. These models typically generate awareness messages at periodic time intervals (100ms to 1s) or with a constant message size (200-400 bytes). These simplified models are used e.g. in [1] with IEEE 802.11p, [2][3] with LTE-V2X, and [4] for comparing the performance of LTE-V2X and DSRC. 3GPP recommended during the LTE-V2X standardization process a traffic model with two message sizes and a fixed time interval between CAMs [5]. An aperiodic traffic model was later introduced in [6], but the model is not compliant with ETSI rules for the generation of CAM messages [7]. These rules specify when vehicles should generate CAMs, and what should be their content. [8] experimentally demonstrated that current standards create CAMs with different time intervals and variable size. This was observed under urban, sub-urban and highway scenarios using commercial and standard-compliant V2X devices. These devices implemented different Facilities layer profiles and were embedded on vehicles of two OEMs. The statistics reported in [8] show significant differences between the collected traces and the CAM messages generated with the simplified traffic models [5][6]. These differences can significantly impact the conclusions reached in studies based on the transmission and reception of awareness messages.

In this context, this paper presents the first set of empirical models to realistically generate CAMs in vehicular networks. The models are created using the real traces presented in [8]. These traces were collected by Volkswagen and Renault in real urban, suburban and highway scenarios. The derived models are based on $m$th order Markov sources. They model the time interval between CAMs (referred to as CAM time interval or CAM generation interval) and the size of CAMs. We present two sets of models. The first one jointly models the size of CAMs and the time interval between CAMs, and is capable to accurately reflect the existing correlation between these two variables. The second set separately models the two variables. These simpler (and less accurate) models have been produced for simulation purposes. The models are validated against the empirical traces reported in [8]. The models can be easily integrated into any network simulator, and are openly provided to the community in [9].

## II. Cooperative Awareness Messages

CAM messages are generated at the Facilities layer of the ETSI ITS Communications Architecture. ETSI defines in [7] the format of CAMs and the CAM generation rules. The format and generation rules are applicable regardless of the technology used for the access layer. ETSI rules specify that CAMs should be generated every 100ms to 1s. A vehicle should generate a new CAM if any of the following triggering conditions is satisfied:
- The distance between the current position of the vehicle and the position included in its previous CAM exceeds 4 m.
- The absolute difference between the current speed of the vehicle and the speed included in its previous CAM exceeds 0.5 m/s.
- The absolute difference between the current heading of the vehicle and the heading included in its previous CAM exceeds 4°.
- The time elapsed since the last CAM was generated is equal to or higher than 1 s.

A vehicle checks these conditions every *T_CheckCamGen*≤100ms, i.e. at least 10 times per second. The time interval between CAMs is

Copyright (c) 2020 IEEE. Personal use of this material is permitted. However, permission to use this material for any other purposes must be obtained from the IEEE by sending a request to pubs-permissions@ieee.org.

This work was supported in part by the Spanish Ministry of Economy and Competitiveness and FEDER funds under project TEC2017-88612-R, and by the TransAID project under the Horizon2020 Framework Programme, Grant Agreement no. 723390.

Rafael Molina-Masegosa, Miguel Sepulcre and Javier Gozalvez are with the Universidad Miguel Hernandez de Elche (UMH), Spain. E-mail: rafael.molinam@umh.es, msepulcre@umh.es, and j.gozalvez@umh.es.

Friedbert Berens is with FBConsulting Sarl in Wasserbillig, Luxembourg, e-mail: friedbert.berens@me.com. Vincent Martinez is with NXP semiconductors France. Email: Vincent.martinez@nxp.com.

then variable and a multiple of *T_CheckCamGen*. The measurements reported in [8] show that it is unlikely that the time between consecutive CAMs is constant for more than 3 CAMs (except when the vehicle is stopped).

A CAM message includes one ITS PDU header and multiple mandatory or optional containers [7]. The header includes data elements (DE) such as the protocol version, the message type and the ID of the vehicle or RSU (Road Side Unit) that transmits the CAM. Each container includes a series of optional and mandatory DEs:

- The basic container is mandatory and includes information of the transmitting vehicle (e.g. the type of vehicle or its position).
- The high frequency container is mandatory and contains highly dynamic information of the transmitting vehicle (e.g. its acceleration, heading or speed).
- The low frequency container is optional and contains static and dynamic information of the transmitting vehicle (e.g. the status of the exterior lights and the vehicle's path history).
- The special vehicle container is optional and is transmitted by specific vehicles such as public transport, emergency vehicles or vehicles transporting dangerous goods.

The size of CAMs depends on the optional containers and the DEs included. The ITS PDU header and the basic container are mandatory and have a fixed size. The high frequency container is mandatory. However, 7 of its 16 DEs are optional. The size of this container is hence variable, and can depend on the manufacturer and the context conditions of the vehicle [8]. The low frequency container is optional and is normally transmitted less frequently than the high frequency container. It has three mandatory DEs including the *PathHistory*. This DE describes the path that a vehicle has followed. The description can use between 0 and 40 path entries, so the size of *PathHistory* is not fixed. The number of path entries depends on the driving conditions and the implementation [8]. For example, the Car-to-Car Communication Consortium (C2C-CC) profile 1.3 [8] establishes that *PathHistory* should cover 200-500 meters of history of a vehicle. However, other implementations like SCOOP release 1.2 [8] propose up to 40 points. Security also has an impact on the amount of data that is finally transmitted. Security certificates might be attached to a CAM before transmission. The certificate is attached whenever a new neighboring vehicle is detected or once per second. The certificate can also be sent on-demand, for example, whenever requested by a RSU. The size of security certificates usually varies between 100 and 150 bytes [8]. Considering the security certificates and the optional containers and DEs, the size of CAMs can vary between 200 and 800 bytes. These variations are significant and should be taken into account to accurately estimate the V2X performance.

## III. EMPIRICAL CAM TRACES

The models presented in this study have been derived using CAM traces obtained by Volkswagen and Renault in test drives described in [8]. The traces have been collected in urban, suburban and highway scenarios under normal road traffic conditions. All traces were generated by On Board Units (OBU) embedded in vehicles, so we do not consider CAM messages generated by RSUs. Each OEM conducted test drives on different locations and using commercial ITS-G5 equipment from different vendors [8]. The traces include (among other) the size and time at which each CAM is generated. This section uses the Volkswagen traces to discuss the major trends observed. Similar trends have been observed with the Renault traces. Differences between both OEMs are highlighted when appropriate.

Fig. 1 depicts the PDF (Probability Density Function) of the size of the CAMs generated by Volkswagen in highway scenarios. The

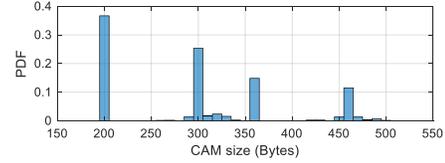

Fig. 1. PDF of the size of CAMs (Volkswagen, highway). Bin size of 10B.

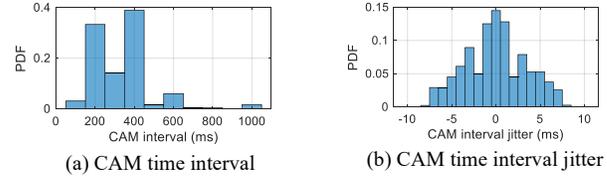

(a) CAM time interval   (b) CAM time interval jitter

Fig. 2. PDFs of the time interval between CAMs and its jitter (Volkswagen, Highway).

figure shows that the size of CAMs is not constant, and that certain values are more probable. In particular, most CAMs have a size of approximately {200, 300, 360, 455} bytes. We conducted a similar analysis with the Renault traces, and found that the most probable CAM sizes for Renault are approximately {200, 330, 480, 600, 800} bytes. The differences observed between both OEMs are due to the different profiles used at the Facilities layer [8]. Similar trends and CAM sizes are observed under urban and sub-urban scenarios for both OEMs.

Fig. 2a plots the PDF of the time interval between CAMs for the highway traces collected by Volkswagen. The results are shown with a time resolution of 100ms since the parameter *T_CheckCamGen* was configured equal to 100ms during the test drives. The CAM time interval is then always a multiple of 100ms. Fig. 2a clearly shows that CAM messages are not generated periodically. The same trend has been observed for the Renault traces and the other scenarios. Fig. 2b shows that there is certain jitter (approximately between -10ms and 10ms). This results in that the CAM time interval is not exactly a multiple of 100ms. Fig. 2b depicts the PDF of the jitter observed in the Volkswagen highway traces. A similar jitter has been observed in all the traces reported in [8]. The jitter can be due to several factors including the time needed to process and encode the CAMs, and the time spent in executing other tasks on the hardware.

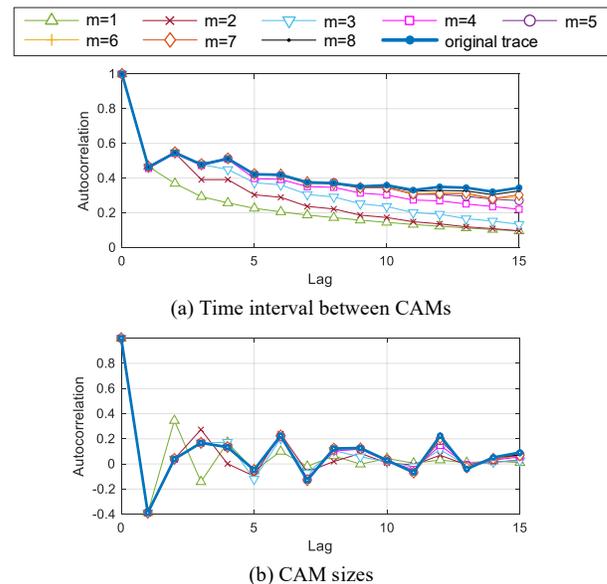

(a) Time interval between CAMs

(b) CAM sizes

Fig. 3. Autocorrelation in the original traces and in traces generated with the proposed models (Volkswagen highway scenario).

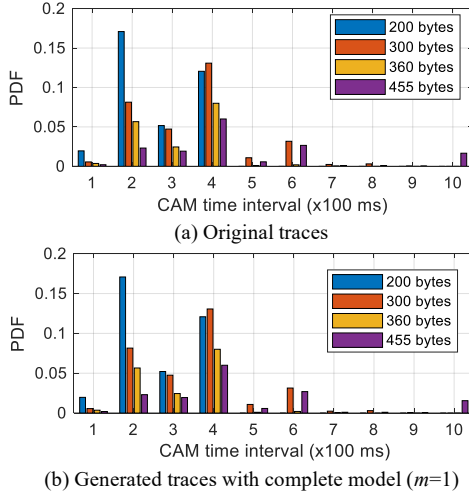

Fig. 4. Joint PDF of the time interval between CAMs and their size (Volkswagen, Highway).

An analysis of the traces collected by Volkswagen and Renault showed that the generation of consecutive CAMs is correlated. This correlation is visible in Fig. 3 that plots the autocorrelation of the CAMs' size and time interval collected in the original traces. This correlation results from the time correlation of the vehicular context and mobility conditions that affect the CAM triggering conditions and the content of the CAM containers. We have also observed in the traces that there is a certain correlation between the size of CAMs and the time interval between CAMs. In particular, we have observed that the probability of generating a CAM with a given size depends on the current time interval. This is observed in Fig. 4a. that plots the joint PDF of the time interval between CAMs and their size for the Volkswagen highway traces. Fig. 4 only considers the four most probable CAM sizes following Fig. 1. Fig. 4a shows, for example, that there is a high probability of generating a CAM of around 200 bytes when the time interval between CAMs is 200ms. However, the probability of generating such CAM decreases with the time interval.

## IV. MODELS

This section presents the models derived from the CAM traces reported in [8] and analyzed in the previous section. They model the size of CAMs and the time interval between CAMs. We present two different sets of models. The first one (*complete models*) jointly models the size of CAMs and the time interval to the next CAM. This approach produces the most accurate models since they account for the correlation between the two variables (Fig. 4a). We have produced a second set of models (*separate models*) where the time interval between CAMs and their size are modelled separately. These models do not capture the correlation between the two variables but have been produced for simulation purposes as an option with lower computational cost for studies where the correlation is not relevant. All models presented in this section are openly available in [9] where we also provide Matlab scripts to facilitate their use.

### A. Complete models

We have produced different models for each OEM since they produce CAMs of different size (Fig. 1). We have created a model for each scenario (urban, suburban and highway) and OEM. We have also produced a *universal* model for each OEM merging the traces from all the scenarios. Each model is a $m$th order Markov source where each symbol in the source alphabet is represented by a CAM with a given size and time interval to the previous CAM. The probability of generating a CAM with a certain size and time interval depends on the $m$ preceding CAMs following Fig. 3. The models are then able to capture the correlation between the size and time intervals of $m$ consecutive CAMs. Each Markov source is specified by the source alphabet $A=\{a_1, a_2, …, a_{|A|}\}$ and a set of conditional probabilities:

$$P(a(t) \mid a(t-1), a(t-2), …, a(t-m)) \quad (1)$$

where $a(t)$ represents the next symbol (at time instant $t$) that depends on the previous $m$ symbols.

The number of symbols in the source alphabet is a function of the number of possible CAM sizes and values of the time interval between CAMs. We define $S=\{s_1, s_2,…,s_{|S|}\}$ as the set of possible CAM sizes. $|S|$ is the cardinality of $S$. We define $G=\{g_1, g_2,…g_{|G|}\}$ as the set of possible time intervals between CAMs. $|G|$ is the cardinality of $G$. The set of symbols in the source alphabet is the cartesian product of $S$ and $G$, i.e. $A = S \times G$. The number of symbols is equal to $|A| = |S| \cdot |G|$. A source symbol $a_n \in A$ is associated to $s_i$ and $g_j$ where:

$$i = ((n-1) \% |S|) + 1; \quad 1 \le n \le |A| \quad (2)$$

$$j = \left\lfloor \frac{n-1}{|S|} \right\rfloor + 1; \quad 1 \le n \le |A| \quad (3)$$

and inversely:

$$n = (j-1) \cdot |S| + i; \quad 1 \le i \le |S| \text{ and } 1 \le j \le |G| \quad (4)$$

The number of conditional probabilities needed to define the proposed $m$th Markov source is $|A|^m$. $|A|^m$ can be very large even for small values of $m$. We define a transition matrix $M$ in order to efficiently store and use the conditional probabilities. Each row in the matrix is defined as follows:

$$(a(t-m) \quad … \quad a(t-2) \quad a(t-1) \quad a(t) \quad P(a(t)|a(t-1),…,a(t-m))) \quad (5)$$

The first $m$ columns represent the previous $m$ symbols (i.e. previous $m$ CAMs). The column $m+1$ represents the next symbol (i.e. the next CAM), and the last column represents the conditional probability that a certain CAM (with a given size and time interval to the previous CAM) is transmitted considering the $m$ previous CAMs. We limit the size of the matrix by removing the rows with a null conditional probability. This reduces the number of rows by several orders of magnitude and significantly improves the tractability of the models.

To compute the conditional probabilities, we parsed the CAM traces and counted the number of times that a certain CAM was generated (with a given size and time interval to the previous CAM) for each set of possible $m$ previous CAMs. For example, the probability that the source generates a symbol $a_n$ for $m=5$ considering that the previous symbols were $a_u$, $a_v$, $a_w$, $a_x$ and $a_y$ is computed as:

$$P(a_n|a_u, a_v, a_w, a_x, a_y) = c_{n,u,v,w,x,y}/r_{u,v,w,x,y} \quad (6)$$

where

$$r_{u,v,w,x,y} = \sum_{n=1}^{|A|} c_{n,u,v,w,x,y} \quad (7)$$

and $c_{n,u,v,w,x,y}$ is the number of times $a_n$ was generated after $a_u$, $a_v$, $a_w$, $a_x$ and $a_y$. The normalization in (6) results in that the sum of the conditional probabilities associated to a symbol $a_n$ for any set of $m$ previous symbols is equal to one.

The jitter observed in Fig. 2b is not directly included in the $m$th order Markov source. Instead, it is computed and added every time a new CAM is generated. The jitter is modelled using a Gaussian distribution with zero mean and standard deviation following Table I.

Following Section III and Fig. 1, all models generated with the traces collected by Volkswagen consider $S=\{200, 300, 360, 455\}$ bytes. Similarly, the models created from the traces collected by Renault use $S=\{200, 330, 480, 600, 800\}$ bytes. We have limited $S$ to these values since other CAM sizes were unlikely. Their inclusion would unnecessarily increase the complexity of the models. We consider for all models that $G=\{100, 200,...,1000\}$ ms. The Markov sources derived using the Volkswagen and Renault traces are then characterized by 40 and 50 symbols, respectively. The complete transition matrices are not shown in this paper for readability reasons given their size. Instead, all matrices are openly provided in [9]. As an example, (8) shows some rows of $M$ ($M$ has in total 1853 rows) for the Volkswagen highway model with $m = 5$.

$$M = \begin{pmatrix} \vdots & \vdots & \vdots & \vdots & \vdots & \vdots & \vdots \\ 13 & 6 & 7 & 14 & 13 & 8 & 1.000 \\ 13 & 6 & 15 & 6 & 5 & 5 & 0.250 \\ 13 & 6 & 15 & 6 & 5 & 14 & 0.750 \\ 13 & 6 & 15 & 10 & 13 & 6 & 1.000 \\ 13 & 6 & 15 & 14 & 5 & 5 & 0.333 \\ 13 & 6 & 15 & 14 & 5 & 16 & 0.667 \\ 13 & 6 & 15 & 14 & 9 & 16 & 1.000 \\ 13 & 6 & 15 & 14 & 13 & 2 & 0.143 \\ 13 & 6 & 15 & 14 & 13 & 8 & 0.143 \\ 13 & 6 & 15 & 14 & 13 & 12 & 0.143 \\ 13 & 6 & 15 & 14 & 13 & 16 & 0.571 \\ \vdots & \vdots & \vdots & \vdots & \vdots & \vdots & \vdots \end{pmatrix} \quad (8)$$

The matrices are provided in [9] in plain text so that they can be easily processed and used. We also provide a Matlab script that describes how CAM messages can be produced using the derived models. The generation of each symbol (or CAM) depends on the previous $m$ symbols. To generate the first symbol, the script randomly selects a preliminary sequence of $m$ symbols from the trace taking into account the probability of each available sequence. To compute these probabilities (also provided in [9]), we parsed the trace and counted the number of occurrences of each sequence. The following steps are then executed to generate a CAM:

1. Select the next symbol of the Markov source $a_n$ considering the previous $m$ symbols and the transition matrix $M$.
2. Identify the size $s_i$ of the next CAM associated to symbol $a_n$ using (2).
3. Identify the time interval $g_i$ at which the next CAM associated to symbol $a_n$ is generated using (3). Generate the jitter and add it to this time interval to compute the exact time at which the next CAM is generated.

The models can be generated for any value of $m$. In [9], we provide models for $m=1$ and $m=5$. The model with $m=1$ reduces the computational cost and can be adequate when the time correlation between symbols is not relevant or has a low impact. Such correlation can be relevant in certain studies, and hence we provide also the model with $m=5$ (i.e. the size and time interval of a CAM depends on the previous 5 CAMs). This model accurately captures the autocorrelation present in the empirical traces (see Section V) and offers a trade-off between complexity and accuracy.

### B. Separate models

TABLE I. JITTER STANDARD DEVIATION (MS)

| Scenario | Volkswagen | Renault |
|---|---|---|
| Urban | 3.235 | 2.817 |
| Suburban | 3.814 | 2.769 |
| Highway | 3.444 | 2.711 |
| Universal | 3.553 | 2.783 |

The complete models generate CAM messages and determine their size and time interval between them. However, users might only be interested to utilize the models to decide the size of CAMs or the time intervals between CAMs. For example, users might utilize traffic mobility simulators (e.g. SUMO) to simulate complex traffic scenarios and compute the exact moment at which CAM messages should be generated following the CAM triggering conditions described in Section II. In this case, users might only be interested in models that determine the size of CAMs. The complete models can be used to determine only the size of CAMs or their time intervals. To do so, the models should be used as explained in section IV.A but discarding step 2 or 3. If step 2 is discarded, the model will only determine the time interval to the next CAM. If step 3 is discarded, it will determine the size of CAMs.

The complete models have a source alphabet $A = S$ x $G$. The resulting transition matrices $M$ are larger than necessary (and hence impact the computational cost) if the complete models are used to only determine the size of CAMs or the time interval to the next CAM. The separate models represent an alternative to reduce the size of $M$ and reduce the computational time. These models independently generate the size of CAM messages or the time intervals between CAMs. They are also modeled using $m$th order Markov sources with $A=S$ and $A=G$, respectively. Each symbol in the source alphabets corresponds to either a possible CAM size or a time interval. The transition matrices are generated following the process described for the complete models. In the case of the complete models, the conditional probability is equal to the probability that a certain CAM with a given size and time interval is transmitted considering the $m$ previous CAMs. In the case of the separate models, it is equal to the probability that a certain CAM with a given size or time interval is transmitted considering the $m$ previous CAMs. The jitter model (including the standard deviations in Table I) is still valid for the separate model that generates the time interval between CAMs. The separate models have smaller alphabets and transition matrices than the complete models and hence represent a lower computational cost alternative. We provide in [9] separate models for $m=1$ and $m=5$.

## V. VALIDATION

This section validates the proposed models. Both complete and separate models have been validated, but we mainly focus in this section on the complete models since they more accurately represent the generation of CAMs in realistic scenarios. For the validation, we generate 5 million CAMs per scenario and OEM using our models. We first then compute and compare the joint PDF of the time intervals between CAMs and CAM sizes obtained with our models and with the real traces. Fig. 4 shows such comparison considering the Volkswagen highway traces. The figure shows the high similarity between the joint PDFs obtained with our model (Fig. 4.b) and with the real traces (Fig. 4.a). The comparison in Fig. 4 is done with $m=1$. A similar joint PDF is obtained with the complete model with $m=5$. A more accurate comparison is done using the KL divergence (or relative entropy) and the total variation distance (or statistical distance) metrics [10]. Both metrics are used to compare PDFs. We denote as $P$ and $Q$ the joint PDFs in Fig. 4 computed with the real traces and the proposed models, respectively. The KL divergence measures the amount of information lost when $Q$ is used to approximate $P$. It is computed as:

$$D_{KL}(P||Q) = \sum_{a \in A} P(a) \cdot Log\left(\frac{P(a)}{Q(a)}\right) \quad (9)$$

where $A$ is the set of possible pairs of CAM sizes and intervals between CAMs (i.e. the source alphabet in our model). The total variation distance between $P$ and $Q$ is the largest possible difference between the probabilities that the two PDFs assign to the same event. It can be expressed as:

$$\delta(P,Q) = \sup_{a \in A} |P(a) - Q(a)| \qquad (10)$$

TABLE II. STATISTICAL COMPARISON

| $m$ | Traces | Scenario | $D_{KL}(P\|\|Q)$ | $\delta(P,Q)$ |
|---|---|---|---|---|
| 1 | Volkswagen | Urban | $2.558 \times 10^{-5}$ | 0.0007 |
| | | Suburban | $8.281 \times 10^{-5}$ | 0.0029 |
| | | Highway | $3.522 \times 10^{-5}$ | 0.0013 |
| | | Universal | $4.621 \times 10^{-5}$ | 0.0015 |
| | Renault | Urban | $2.388 \times 10^{-4}$ | 0.0018 |
| | | Suburban | $3.145 \times 10^{-4}$ | 0.0014 |
| | | Highway | $2.046 \times 10^{-4}$ | 0.0018 |
| | | Universal | $1.603 \times 10^{-4}$ | 0.0015 |
| 5 | Volkswagen | Urban | $2.606 \times 10^{-5}$ | 0.0008 |
| | | Suburban | $7.692 \times 10^{-5}$ | 0.0027 |
| | | Highway | $4.289 \times 10^{-5}$ | 0.0014 |
| | | Universal | $4.548 \times 10^{-5}$ | 0.0015 |
| | Renault | Urban | $2.647 \times 10^{-4}$ | 0.0018 |
| | | Suburban | $3.087 \times 10^{-4}$ | 0.0013 |
| | | Highway | $1.687 \times 10^{-4}$ | 0.0017 |
| | | Universal | $1.911 \times 10^{-4}$ | 0.0015 |

Table II reports the two metrics computed for different scenarios. The results clearly demonstrate the high similarity between the joint PDFs obtained with the real traces and with our complete models.

The separate models do not generate CAM messages with the same accuracy as the complete ones since they do not account for the correlation between the CAM sizes and the time intervals. They can still be relevant for simulation purposes and it is interesting to analyze their accuracy. We first compute the KL divergence and total variation distance metrics when generating the joint PDF of the CAM time intervals and sizes with the separate models (i.e. without the correlation between both variables). In this case, $D_{KL}$ and $\delta$ are equal to 0.1093 and 0.0485 respectively when considering the Volkswagen highway traces and the traces generated with the separate model with $m=1$ model (similar values are obtained with $m=5$). The metrics improve if we compare the PDFs of the CAM sizes (or the PDFs of the CAM time intervals) obtained with our separate models with that obtained with the real traces. In this case, $D_{KL}$ and $\delta$ are equal to $5.3192 \times 10^{-6}$ and 0.001 respectively when considering the PDFs of the CAM sizes and the Volkswagen highway scenario. $D_{KL}$ and $\delta$ are equal to $1.5549 \times 10^{-5}$ and 0.0021 respectively when considering the PDFs of the CAM time intervals. Similar trends have been observed for all the scenarios and OEMs.

Fig. 3 compares the autocorrelation of the CAMs' size and time interval observed in the original traces and in traces generated with the complete and separate models. The performance obtained with the complete model is shown in Fig. 3 with different values of $m$. The figure shows that the complete model with $m=1$ does not capture accurately the time correlation present in the original traces. This correlation is though accurately modelled when $m=5$. In this case, the absolute difference of the autocorrelation observed in the generated and original traces is less than 0.1 for sequences of up to 15 consecutive symbols. The accuracy gained with higher values of $m$ is not significant and does not justify the higher computational cost resulting from larger transition matrices $M$ when $m$ increases. The proposed model with $m=5$ offers then an adequate trade-off between computational cost and accuracy, and this is why we published in [9] models with $m=1$ and $m=5$. The separate models match also the autocorrelation observed in the original traces, but with less accuracy than the complete models. However, the separate models cannot model the correlation between the size of CAMs and the time interval between CAMs observed in Fig. 4. This is visible in Fig. 5 that plots the cross-correlation between the CAMs' size and time intervals for the original traces and the traces generated with the complete and separate models. Fig. 5 shows that only the complete model with $m=5$ can accurately capture the cross-correlation present in the original traces. The separate models should then be utilized in studies where the correlation between the size and time interval of CAMs is not critical. It should be noted that although this correlation exists, it is not too high (<0.4, Fig. 5).

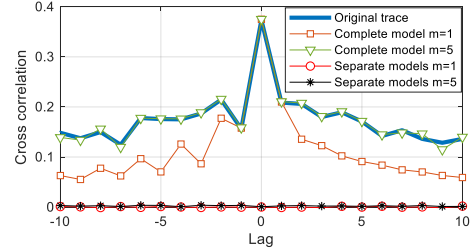

Fig. 5. Cross-correlation in the original traces and in traces generated with the proposed models (Volkswagen highway scenario).

## VI. CONCLUSIONS

This paper proposes a set of models to realistically generate Cooperative Awareness Messages (CAMs) in vehicular networks. To the authors' knowledge, these models are the first publicly available that have been created from real CAM traces collected in test drives using commercial standard-compliant V2X equipment. The models have been derived using traces collected by Volkswagen and Renault in urban, suburban and highway test drives under normal road traffic conditions. The proposed models are based on $m$th order Markov sources. They model the size of CAMs and the time interval between CAMs. This study presents two sets of models. The first one jointly models the size of CAMs and the time intervals, and is hence capable to account for the correlation between these two variables. A simpler set of models is also presented for simulation purposes. These models separately (and independently) model the size of CAMs and the time intervals. All models can be easily integrated into any network simulator and are openly provided to the community in [9]. The availability of realistic traffic models is necessary for an accurate evaluation of vehicular networks.